\documentclass[a4paper,fleqn]{cas-dc}
\usepackage{amsmath}
\usepackage{mathtools}

\usepackage[utf8]{inputenc}
\usepackage[numbers]{natbib}
\usepackage[linesnumbered,ruled]{algorithm2e}

\def\tsc#1{\csdef{#1}{\textsc{\lowercase{#1}}\xspace}}
\tsc{WGM}
\tsc{QE}
\tsc{EP}
\tsc{PMS}
\tsc{BEC}
\tsc{DE}
\begin{document}
\let\WriteBookmarks\relax
\def\floatpagepagefraction{1}
\def\textpagefraction{.001}
\shorttitle{Leveraging Artificial Intelligence to Analyze Citizens' Opinions on Urban Green Space}
\shortauthors{M. Ghahramani et~al.}

\title [mode = title]{Leveraging Artificial Intelligence to Analyze Citizens' Opinions on Urban Green Space}                      
% \tnotemark[1,2]

\tnotetext[1]{This paper is the result of a research project funded by Connecting Nature (Grant Agreement No. 730222) under the European Community's Framework Program Horizon 2020 and the Netherlands Fulbright Association through the Fulbright U.S. Doctoral Student Program. This research was made possible by the support of the Senseable City Laboratory at the Massachusetts Institute of Technology, which hosted Nadina Galle during her fieldwork from September 2019 to March 2020.}

\author[1]{Mohammadhossein Ghahramani}
\ead{sepehr.ghahramani@ucd.ie}

\author[1,2]{Nadina J. Galle}
\ead{nadina.galle@ucd.ie}

\author[2]{Fábio Duarte}
\ead{fduarte@mit.edu}

\author[2]{Carlo Ratti}
\ead{ratti@mit.edu}

\author[1]{Francesco Pilla}
\ead{francesco.pilla@ucd.ie}
% \ead[URL]{www.sayahna.org}

% \ead[URL]{www.stmdocs.in}

\address[1]{Spatial Dynamics Lab, University College Dublin, Ireland}
\address[2]{Senseable City Laboratory, Massachusetts Institute of Technology, USA}
% \cortext[cor1]{Corresponding author}

\begin{abstract}
Continued population growth and urbanization is shifting research to consider the quality of urban green space over the quantity of these parks, woods, and wetlands. The quality of urban green space has been hitherto measured by expert assessments, including in-situ observations, surveys, and remote sensing analyses. Location data platforms, such as TripAdvisor, can provide people's opinion on many destinations and experiences, including UGS. This paper leverages Artificial Intelligence techniques for opinion mining and text classification using such platform's reviews as a novel approach to urban green space quality assessments. Natural Language Processing is used to analyze contextual information given supervised scores of words by implementing computational analysis. Such an application can support local authorities and stakeholders in their understanding of—and justification for—future investments in urban green space.
\end{abstract}

\begin{keywords}
Sentiment Analysis \sep Supervised Learning \sep Natural Language Processing  \sep Imbalanced Classification
\end{keywords}

\maketitle

\section{Introduction}
Urban Green Space (UGS) such as parks, woods, and wetlands represent a fundamental component of any urban ecosystem. In addition to the many ecological, economic, and psychological benefits, since the 1800s, UGS have been recognized for their ability to offer refuge from pervasive air pollution, and congestion \cite{Barton2017importance,Douglas2017Green,Swanwick2003Nature}. Today, the ecological benefits of green in the city are well-documented, but there is also a growing body of evidence of its positive impact on human health and well-being \cite{Swanwick2003Nature}. Green space offers citizens more opportunities for social contact and stress relief – whether impromptu or planned \cite{Zhang2017Quality}. Studies show UGS should be of critical importance to public mental health, especially from an urban planning perspective \cite{Grahn2010relation,Roe2018Green}. 

For many citizens, UGS has become an extension of, or in many cases the replacement of, the traditional backyard, meaning more people are sharing less green space. Despite appeals for green space's place in the city's master plans and worldwide urban population growth, UGS has decreased in several cities  \cite{Fuller2009Green,Giezen2018Using,Haaland2015Challenges,Nowak2018Declining}. Lucrative urban development and construction are often to blame for its demise. To meet demand, studies have suggested the quality of green space significantly contributes to neighborhood satisfaction and well-being, independent of the quantity of green space \cite{Zhang2017Quality}. How to measure the quality of UGS has been hotly debated in urban forestry and planning fields, with several attempts made to streamline and standardize quality assessments of UGS \cite{Cohen2009Recreational,Daniels2018Assessment,Gidlow2012Development,Girling2005Skinny}. However, with current methods relying on expert assessments, some warn it discredits the experience of local users; who are likely more qualified to assess their own UGS than outside experts \cite{Hur2010Neighborhood}. 

The definition of quality UGS is still contested, and their role remains undervalued. Measuring UGS quality is also a tedious process; observational techniques are often criticized as they require extensive repeat measurements at the same location, incurring large time and cost expenditures \cite{Roberts2017Geo}. Even when data is collected, it is quickly outdated, leaving progress out of reach \cite{Plunz2019Twitter}.

How can a city ensure it provides safe, inclusive, high-quality UGS for all? Emerging technologies are gaining traction as a way to gain up-to-date information on—and engage local users in—the planning and improvement of UGS \cite{Galle2019Internet,Nitoslawski2018Smarter,Ghahramani2020Urban}. There are several quantitative approaches to analyzing UGS such as drones \cite{Nasi2018Remote}, satellite imagery \cite{Fang2020Street}, and Google Street View images \cite{Seiferling2017Green}, but there is a need for reproducible qualitative analysis. One such technology, Natural Language Processing (NLP), combines computer science and linguistics to understand the language in a piece of text. One of its applications, Sentiment Analysis (SA), can extract and categorize positive, negative, or neutral sentiment from a chunk of text. While NLP can work on any written text, performing SA on georeferenced crowd-sourced data sources such as TripAdvisor, Twitter, Yelp, Booking.com, and Airbnb have shown particular promise \cite{Chen2017Delineating,Plunz2019Twitter,Van2018Aesthetic}. Applications range from understanding consumers' attitudes toward their products to the socioeconomic status of communities to hospitality organizations' performance \cite{Kourtit2019Cultural,Ma2018Sentiment,Rahimi2018Geography}. It has been suggested that in most of these applications, sentiment analysis should become a complementary tool for quality assessment and evaluation \cite{Garcia2010}.

TripAdvisor is a particularly popular platform with a rich and publicly accessible database on attractions, destinations, and landmarks, including UGS \cite{Kourtit2019Cultural}. While relevant to this research, studies of the demographic makeup of TripAdvisor are limited. Some groups are likely over- and/or underrepresented on TripAdvisor, but it is still advantageous over population-based surveys, a costly and tedious method to acquire representative population samples. TripAdvisor offers a viable, complementary method to harvest local opinion and feedback on UGS. 

This paper presents a novel NLP application using TripAdvisor to assess the quality of UGS. The corpus collected were TripAdvisor reviews of St. Stephen's Green, the most popular public park in Dublin (Ireland). St. Stephen's Green, in the middle of Dublin's city center, is a 10-hectare park with over eight million visitors on an annual basis. The park has well-maintained facilities on the grounds, including over 3 km of walking paths and public restrooms inside the park. 

Experimental, computational analyses were implemented via two scenarios, and different phases have been included to address identifying the sentiment expressed in reviews. The proposed method allows the extraction and interpretation of sentiment with minimal human effort by applying Artificial Intelligence (AI) and Machine Learning (ML) algorithms. The contributions of this work are as follows:

 \begin{itemize}
  \item We present a novel application of NLP and text mining using TripAdvisor to assess the quality of UGS.
  \item We present how a self-contained sentiment analysis model can be implemented to evaluate people's attitudes toward various entities given a class imbalanced issue.
 \end{itemize}

The paper is organised as follows: some related work in the field of sentiment analysis and opinion mining is presented in Section \ref{RelatedWork}; the proposed approach with its associated discussions is presented in Section \ref{Methods}; Section \ref{Results} shows the experimental results; Section \ref{Discussion} details the discussion; and Section \ref{conclusion} concludes the paper.

\section{Related Work}\label{RelatedWork}
The value of UGS remains underestimated due to a lack of information about what quality green space entails and how existing spaces within the city score on important social-quality parameters. Measuring the quality of UGS is a tedious process. Observational approaches are often criticized as they require extensive repeat measurements at the same location, incurring large time and cost expenditures \cite{Cohen2009Recreational,Roberts2017Geo}. Even when data is collected, it is quickly outdated, leaving progress out of reach.

\subsection{Web-based civic participation platforms}
A recent improvement is web-based civic participation platforms. In an effort to gain insights into how people perceive a park’s quality, several cities have released apps. Amsterdam recently launched “MyPark”, an app that asks the user questions about specific areas of a specific park. Once the results are analyzed, the feedback is incorporated into a park redesign to better meet local user needs. 

FixMyStreet is another example. The map-based app acts as a liaison between residents and their local authority on problems such as potholes and broken street lights needing their attention. The app was launched in the United Kingdom (UK) and has proliferated across the country. FixMyStreet also has an open-source platform that helps people run similar websites all over the world. Although mainly used for reporting common street problems, the app could also be used to highlight issues facing urban parks and woodlands.

\subsection{Microblogging platforms}
Microblogging platforms challenge users to summarise their thoughts in a limited amount of words. Twitter, arguably the world’s microblogging pioneer, allows 280 characters per post (or “tweet”), a recent upgrade from the iconic 140 characters they used to enforce. In her inspiring paper \cite{Roberts2017Geo}, Roberts proposes the use of crowdsourced, geotagged social media data, such as tweets, to inform how, when, and why people use UGS. This method overcomes some issues with previous approaches, such as report based methods, which are difficult to validate, and observational methods, which require multiple observations over different days and seasons to ensure reliability \cite{Roberts2017Geo}. It can even be used to derive seasonal variation in physical activity in UGS \cite{Roberts2017Geo}. Crowdsourcing data from Twitter offers an alternative as it is publicly available and instantly accessible, incurring no additional time or costs.

Yet, both web-based civic participation platforms and social media data face limitations. FixMyStreet, with over 12,000 reports sent to UK councils every month, has much less usability than Twitter, which recorded 17 million monthly active British users in the first quarter of 2018. It is unlikely any significant amount of these tweets are actually about issues on the street, but it is a much broader data source. 

There are also socio-demographic concerns regarding the user base of both FixMyStreet and Twitter. In 2017, \cite{Pak2017FixMyStreet} analyzed over 30,000 FixMyStreet reports, compared them to a range of socio-demographic indicators, and revealed crowdsourced civic participation platforms tend to marginalize low-income and ethnically diverse communities. 

In the same way, the elderly population, who show lower levels of engagement with these forms of technology, are disregarded explicitly in such research \cite{Chen2017Engaging}. This is especially concerning as urban parks are supposed to be a shared public space for all ages. Roberts also reports Twitter data lacks demographic information about Twitter users such as their age, occupation, or ethnicity \cite{Roberts2017Geo}. Although not crucial to determine opinions, these parameters are useful for further examination of where particular attitudes may originate. Evidently, there is a need for inclusive, unrestricted, unbiased, and freely-solicited opinions about UGS.

\subsection{TripAdvisor}
NLP is used to understand the language in a piece of text and reveal the sentiment behind it. The method combines computer science and linguistics. In recent years, the popularity of virtual assistants like Siri, Alexa, and Google Home has accelerated the demand for voice user interfaces. And, as such, increased research on how computers understand speech and speak themselves. 

NLP can also work on written text, like user-generated reviews on the world’s largest travel website, TripAdvisor. The open, online community reaches 390 million unique visitors each month and lists 465 million reviews and opinions about more than seven million accommodations, restaurants, and attractions in 49 markets worldwide. TripAdvisor is a treasure trove of sentiment. When writing a review, a reviewer is prompted to describe their first-hand experience causing both tourists and locals to flock to TripAdvisor to express their opinions. Whereas Twitter offers a platform for sharing an occasional opinion, TripAdvisor explicitly asks for the sentiment.

The overall user base of both TripAdvisor and Twitter is still poorly understood. In 2007, Gretzel \cite{GretzelU} found frequent travel review readers tend to be younger, have slightly higher incomes, and are more likely to contribute to online content. They are also more likely to post reviews themselves. Thus, one could assume the reviewers share a similar demographic to the reader, at least in 2007. 

Gender differences can also play a role. Blumenthal \cite{BlumenthalM2014} found little to no gender differences amongst reviewers on TripAdvisor in 2014. Twitter, on the other hand, did exhibit gender differences. According to Statista, an online statistics, market research, and business intelligence portal, during a 2018 study period, 42.8 percent of global Twitter users were female, and 57.2 percent were male (Global Twitter User Distribution by Gender 2018 | Statistic, n.d.).

The use of Twitter tends to drop as age increases. In the United States, those under 50, especially those 18-29, are most likely to use Twitter. And only 6 percent of Twitter users constitute the 65+ age group. TripAdvisor's TripBarometer report showed a slightly more even distribution of the travel site’s user base. 

Research to validate these demographic claims is limited, and studies comparing TripAdvisor with Twitter’s user base are non-existent. Although some groups remain over- and/or underrepresented on TripAdvisor, it is advantageous over Twitter as it generally covers a broader demographic spectrum. In fact, the only known method to encompass a general population is population-based surveys, where an experiment is administered to a representative population sample. However, this process is costly and lengthy, and as such, TripAdvisor offers a viable, complementary method to harvest local opinion and feedback on UGS. 

So far, sentiment analysis using TripAdvisor as a data source has only been applied in the hospitality and tourism sectors. Here, shallow NLP techniques are applied to extract sentiment \cite{Garcia2010} automatically. These simple expressions, which are derived from the reviews, can be used to evaluate the quality of hotels or restaurants. García-Barriocanal’s preliminary study \cite{Garcia2010} was able to identify emotion types with reasonable effectiveness and suggested sentiment analysis using TripAdvisor reviews should become a complementary tool for hospitality evaluation.

\subsection{Supervised Text Classification}
Sentiment classification is an example of a supervised machine learning task, a process of assigning text documents into two or more predefined classes. In this process, an algorithm takes any observation (text document) as input and assigns a label from the class labels \cite{Ghahramani2020AI,Ghahramani2020Spatial}. Different data-driven supervised approaches have been used to deal with such a classification problem. Sentiment classification has raised much attention in recent years and has undergone many changes. Generally speaking, three techniques can be used to construct a sentiment lexicon, i.e., dictionary-based, corpus-based, and hybrid methods. Dictionary-based methods use word matching based on the lexicon. However, since sentiment words in the lexicon might be difficult to recognize, many texts cannot be analyzed by utilizing such classifiers. Corpus-based methods use labeled data, and lexicons are not effectively taken into account in such approaches. To alleviate the discussed shortcomings, a hybrid approach (i.e., a combination of machine learning methods and lexicons) can help improve the sentiment classification performance. Since the text classification problem is a supervised learning task in which the class observations is predicted based on some feature values, a wide range of ML algorithms (e.g., Support Vector Machine (SVM) \cite{FITRI2019765,ALAMRANI2018511}, Naive Bayes (NB) \cite{FITRI2019765,RAKHMANOV2020194}, decision tree \cite{FITRI2019765}, random forest \cite{FITRI2019765,ALAMRANI2018511}, logistic regression, and neural networks \cite{Jin2020Learning,YAO2020101522,DONG2020103418,COLONRUIZ2020103539}) can be incorporated.

As explained, in this work, people reviews as to UGS are taken into account. These texts are unstructured; thus, manually analyzing them can be tedious and time-consuming. In this type of data mining, people's opinions, sentiments, and attitudes are analyzed. The main objective is to computerize the process of reading reviews and evaluate them. It should be mentioned that the most crucial task in sentiment analysis is the pre-processing phase, including different operations. Due to differences in data characteristics, these tasks might differ from one sentiment analysis approach to another. Because of the complexity of feature dependency, ML methods may achieve different results. Given this work's characteristics, we aim to propose an appropriate approach to deal with various issues to be explained next. It is worth mentioning that a self-contained model consisting of multiple phases is implemented in this paper. In the last stage of the model (the classification phase), different ML algorithms are tested, and their corresponding results are compared.

\section{Method}\label{Methods}
\begin{figure}
  \includegraphics[width=\linewidth]{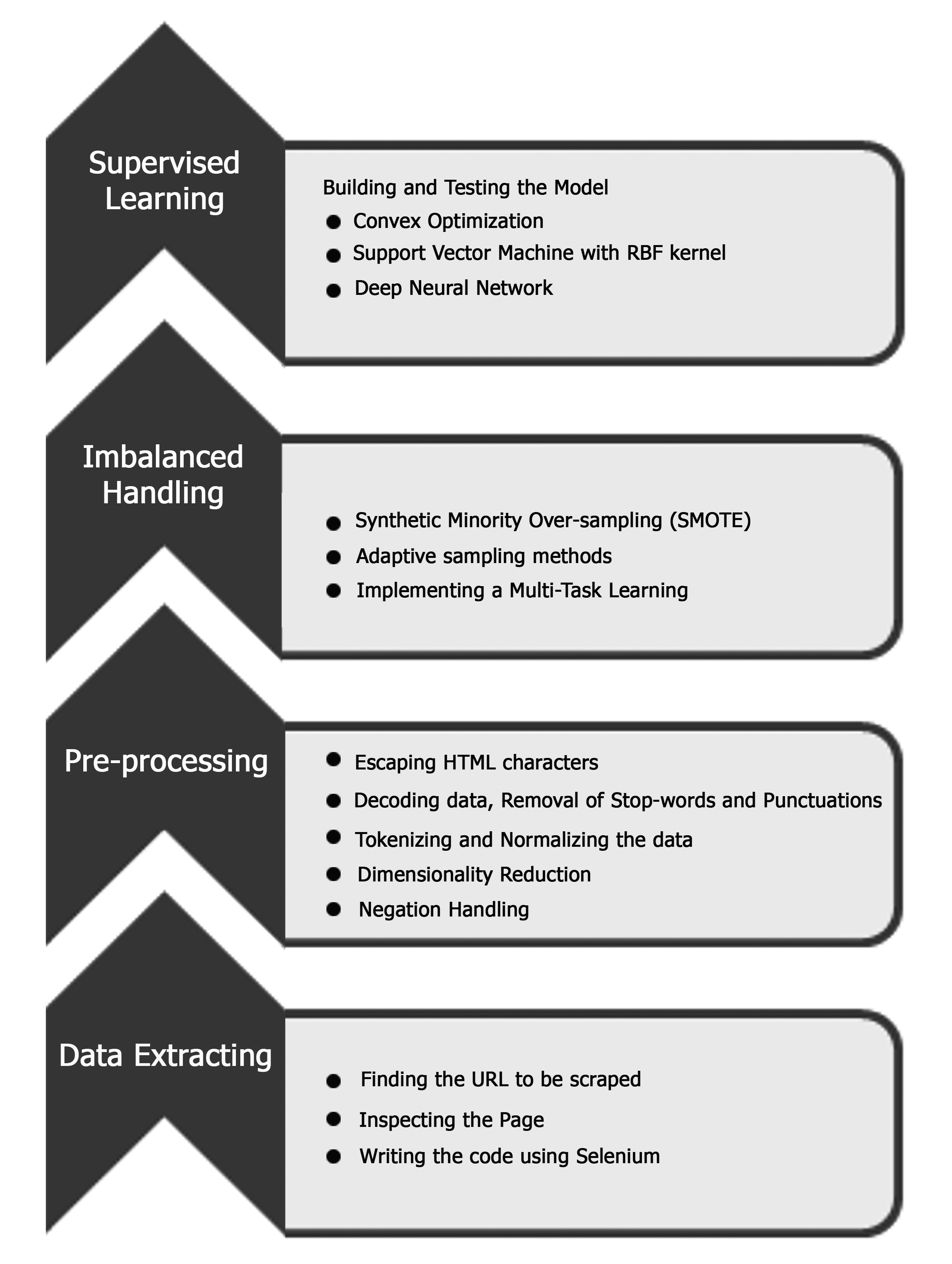}
  \caption{Four phases of the sentiment analysis model used in this work.}
  \label{Model}
\end{figure}

There have been different approaches to perform sentiment analysis. However, choosing a proper method is highly related to the nature of a given work. This paper analyzes people's opinions and sentiments to identify different positive and negative polarities on urban green space. Different from book articles and news reports, review texts are often short and ambiguous. Various models, i.e., fully supervised and semi-supervised methods, have been considered to analyze review comments. The methods in the former category use manually labeled data. Their approach is very time-consuming to create lexicons manually. Some specific supervised methods have been introduced to train sentiment classifiers on emoticons and hashtags. Because of such shortcomings, a semi-supervised model has been considered in this paper.
It should be noted that the classification phase of the model is based on a supervised technique, while an unsupervised method is used in the pre-processing phase of the model. Most of the concerns related to opinion mining and sentiment analysis of reviews can be addressed by implementing effective pre-processing techniques. However, there are no effective pre-processing methods for all datasets and algorithms. For instance, in this work, we deal with an imbalanced classification issue since most comments in the dataset used are positive. A multi-layer approach consisting of different phases (i.e., web scrapping, data cleanings, imbalanced classification, and supervised ML) is implemented to address all concerns. Fig. \ref{Model} illustrates different phases of the proposed model. The following sections present all details regarding each stage of the model. 

\subsection{Data Extracting}
TripAdvisor reviews for St. Stephen's Green (Dublin, Ireland) were scarped using Selenium and Python. The pseudocode is presented in Algorithm \ref{scrapping}. The reviews were subsequently processed to focus on English texts, because of the mass availability of English language text analysis tools and dictionaries. The reviews were collected from the period of May, 2006 to November, 2020, for a total of 16,613 reviews; in contrast, Dublin's second most popular park had 4,753 reviews for this time period. Of the St. Stephen's Green reviewers, 5,622 were from the United States, 3491 were from Ireland, 2,835 from the United Kingdom (UK), 796 from Canada, 483 from Australia, 105 from Germany, 92 from the Netherlands, 91 from South Africa, 48 from New Zealand, 43 from Denmark, 39 from Greece, 32 from United Arab Emirates, and 30 from China. The remaining 2,906 reviewers were from other countries, had misspelled their country, or left their location blank.

\begin{algorithm*}
\SetAlgoLined

    \SetKwInOut{Input}{Input}
    \SetKwInOut{Output}{Output}
   % \Input{ $W{ij}$ Matrix (the contiguity matrix)}
    \Input{$CSV file \gets$ [Score, Date, Title, Review]; $driver \gets \text{webdriver.Chrome()}$;
     $ Fn \gets \text{Selenium.find\_element\_by\_xpath}$;
     $URL \gets \text{https://www.tripadvisor.ie/*.st\_stephens\_green}$\\
    }
    
    \Output{Review Comments}
    \# \enspace \textit{Exception Function} \\
    {try:}\\
    \enspace \text{            }\text{            }{        driver.find\_element\_by\_xpath(xpath)}\\
    {except NoSuchElementException:}\\
    \enspace \text{            }\text{            }{        return False}\\
    {return True}\\

        {n = number of web pages}\;
        \# \enspace \textit{HTML elements} \\
        {$e_0$= 'taLnk  ulBlueLinks'};
        {$e_1$= 'ui\_bubble\_rating bubble\_'}; 
        {$e_2$= '\_34Xs-BQm'};
        {$e_3$= 'glasR4aX'};
        {$e_4$= 'IRsGHoPm'};
        {$e_5$= 'Dq9MAugU T870kzTX LnVzGwUB'};

        \For{$i \gets 1,2, .... , n$}{
        \If{$Exception Function(''//span[@class=e_0]'')$}
    	{ $driver.Fn(''//span[@class=e_0]'').click();$}
		
    $df \gets driver.Fn(''//div[@class=e_5]'')$;\\                    
     $num \gets  len(df)$;\\              
       \For{$j \gets 1, 2,..., num$}{
    	{ $Score = df[j].Fn(''.//span[contains(@class, e_1)]'').get\_attribute(''class'').split(''\_'')[3]);$}\\
    	{ $Date = df[j].Fn(''.//span[@class=e_2]'').get\_attribute(''title'');$}\\
    	{ $Title = df[j].Fn(''.//div[@class=e_3]'').text;$}\\
    	{ $Review = df[j].Fn(''.//q[@class=e_4]'').text.replace(''\textbackslash n'', ''{ }'');$}\\

		}   
	}

    {Return: Score, Date, Title, Review.}
    \caption{Pseudo-code for extracting Tripadvisor reviews}
    \label{scrapping}
    \end{algorithm*}

The following review fields were extracted: review-title (written title of review); review-body (written review about the destination); rate-value (1 is the lowest evaluation, 5 is the best); review-location (where a reviewer is from); and review-date (date review was written). Only review-body and rate-value data fields were used in this experiment. This dataset can be considered as a sequences of text, i.e., $D = \{X1 , X2, . . . , Xn\}$ where $X_i$ refers to the \text{$i^{th}$} review. Each review is also labeled as positive or negative, depending on its corresponding rate value.

\subsection{Pre-processing}
As the quality of data affects the analysis, it is essential to employ a data pre-processing procedure. To that end, feature extraction was performed, and a structured set from the reviews is created for the model-training purposes. A dimensionality reduction operation is also considered by applying the Term Frequency-Inverse Document Frequency (TF-IDF) technique \cite{Yahav2019Comments}. These pre-processing steps help us convert unstructured text sequences into a structured feature space. Data cleansing operations were performed, and punctuations and stop words were omitted. To make transformations (removing punctuations, stop words, and other cleansing operations) implemented in this work, libraries from the Natural Language Toolkit (NLTK) were used. This Python library has been written for modeling text and provides various tools for loading and cleaning texts. This library's different functions were used for filtering punctuation, stemming, normalizing, extracting text from HTML, decoding Unicode characters, locating typos, and handling numbers. 

Data normalization techniques (e.g., Stemming and Lemmatisation) were applied, each review was converted to a numeric representation (corpus), and the $n$-grams approach (with two different measures like Word Counts and TF-IDF) was implemented. The former is based on mapping more than just one word (unigrams) onto the corpus. We have also included word counts into our model. To that end, the number of times a given word or a sequence of words appear is counted. The latter, term TF-IDF, is a weighting measure to be used instead of word count representations. This measure is considered to lessen the effect of implicitly common words in the corpus. The weight of a term in a review can be defined as:

    \begin{equation}
w(r,t) = TF(r,t) * log(\frac{n}{df(t)})
  \end{equation}

where $n$ is the number of reviews and $df(t)$ is the number of reviews consisting of the term $t$ in the corpus.

Negation handling could be another challenging task for sentiment classification. However, since we deal with a two-class classification task, such concern can be easily addressed. The model negates the predicted class of observations, as there are only two classes to choose from. Such negation recognition can be a complicated process in cases where there are more than two possible classes. In our case, the negation handling procedure is considered as an Exclusive-OR problem. As far as the negation scope detection is concerned, different negation keywords are defined, and the regular expression-based NegEx method is used. Moreover, the negation implicitly is captured via n-grams.

Although the dataset has been cleaned after performing the explained operations, there is still a considerable concern. The most challenging data pre-processing task in this work is an imbalanced class issue. This procedure is usually regarded as a pre-processing task; however, we consider it as a separate task to be explained next due to its importance in our work.

\subsection{Imbalanced Handling}
The observations that were labeled as negative are relatively rare as compared to the positive class (less than 10\%). Positive and negative labels are determined according to the reviewers' ratings. Should a rating be higher than four, the corresponding review is considered as a positive one; otherwise, it is treated as a negative review (Table 1).

\begin{table*}[ht]
\caption{Examples of positive and negative TripAdvisor reviews about  St. Stephen’s Green (Dublin, Ireland).} % title name of the table
\centering % centering table
\footnotesize
    \begin{tabular}{| c | p{13cm}  | l |}
    \hline
    Bubble\_rating & Review\_body & Label \\ \hline
    5 & Stephens Green is a great place to visit nice walk around the park there also in the summertime music playing you can have a picnic there 
    watch the ducks and the swans in the pond there is also boards giving a bit of history...& positive  \\ \hline
    5 &  Great place to relax in the city. Beautiful gardens and paths to walk around. If you need to just sit a bit this is a great place to do so. & positive \\ \hline
    4  & St. Stephen's Green park is perfect to step away from the hustle and bustle of Dublin. 
    The scenery is beautiful and calming. Then when you are refreshed you just step back into the action of the city.& positive \\\hline
    3  & Only downside is anti-social behaviour. Always somebody hassling for money or asking for a smoke. 
    Wouldn't mind it on my own but with kids it's terrible. Should be better patrolled. Europe's...& negative \\\hline
    1  & While on a visit to Dublin we brought our children to the park. 
    The place is really nice but we were really shocked when we went to the playground. 
    Near the playground entrance there were about 200 teens drinking and causing trouble.& negative \\\hline
    2  & While we were walking across the park, a young man tried to take my husband's laptop. It was zipped inside a shoulder bag. 
    I yelled and this person went away.& negative \\
    \hline
    \end{tabular}
\label{tab:records}
\end{table*}

Hence, we face an imbalanced classification issue. In other words, positive class (the majority) outnumbered negative class (the minority), and both classes do not make up an equal portion of our dataset. The conventional classifiers such as Decision Tree (\cite{Rokach2005Top}) and Logistic Regression (\cite{Pampel}) do not accurately measure model performance when faced with imbalanced datasets. They usually have a bias towards the majority class, and the minority class observations are treated as noise. To handle this issue, an SVM that performs well against highly imbalanced datasets are used to train our model. This classifier is also equipped with a class weight measure to alleviate the situation. Moreover, a separate imbalanced classification phase is embedded into the model. In doing so, different types of algorithms, i.e., multi-task learning \cite{9076160}, adaptive sampling \cite{8785940}, and synthetic oversampling method \cite{8065074}, were integrated and tested. 

Generally speaking, there are two distinctive approaches for handling the mentioned issue: 1) skew-insensitive techniques and 2) re-sampling approaches. The former deals with a class imbalanced problem by assigning a cost measure to the training data. The latter adjusts the original dataset such that a more balanced class distribution is achieved. Re-sampling methods (\cite{8065074}) have become standard approaches and have been dominantly utilized recently. They can be classified into different categories, e.g., sampling strategies, wrapper approaches, and ensemble-based methods. Implementing a proper method is crucial; otherwise, it can be problematic, e.g., data loss and overfitting, and can result in a poor outcome. This phase aims to balance class distribution relatively. As stated, three different techniques have been tested. We have found that a synthetic oversampling algorithm (\cite{8065074}) performs better than the other two methods (i.e., adaptive sampling and multi-task learning). It is worth mentioning that the two other methods used are also computationally expensive. The synthetic oversampling algorithm creates synthetic samples based on the nearest neighbor approach. By implementing the method, \textit{Failure class} instances are synthetically created, and the distribution is more balanced. The procedures are as follow:

\begin{itemize}
  \item Let $A$ be the set of all elements of the minority class. The algorithm detects k-nearest neighbors of all observations ($S \in A$) of this class. In doing so, the Euclidean distance between each observation and other elements is measured.
  \item A sampling rate (e.g., 60\%) is defined based on the imbalanced proportion. Given such a pre-defined rate, 60\% of k-nearest neighbors of each observation in the minority class are randomly selected. Let $A^{\prime}$ be the set of k-nearest neighbors.
   \item For each element in the obtained set ($S^{\prime} \in A^{\prime}$) the following formula is used to create new samples.
    \begin{equation}
S_{new}=S+\alpha * |S-S^{\prime}|
  \end{equation}
where $\alpha$ is a random number between 0 and 1.
\end{itemize}

The pseudo-code of the procedure integrated into the model to handle class imbalanced issue is presented in Algorithm \ref{imbalancedIssue}.

\begin{algorithm}
\SetAlgoLined

    \SetKwInOut{Input}{Input}
    \SetKwInOut{Output}{Output}
   % \Input{ $W{ij}$ Matrix (the contiguity matrix)}
    \Input{$m \gets$ number of minority class observations;\\ $r \gets$ amount of over-sampling (\%);\\
     $ k \gets$ number of neighbors;
    }
    
    \Output{r*m synthetic observations}
    \# \enspace \textit{Exception Function} \\

        {$f$ = len(features)     \# \enspace \textit{number of features}}\;
        {$S \gets$ [ ]     \# \enspace \textit{observations in the minority class}}\;
        {$S^{\prime} \gets$ [ ]     \# \enspace \textit{synthetic observations}}\;
        {$\Gamma \gets$ [ ]}\;
        \For{$i \gets 1, 2,..., m$}{
    	{ $D \gets \text{Compute Euclidean distance};$}\\
    	{ $D.sorted()$ \# \enspace \textit{sort in an ascending order}}\\
    	{ $Index\gets$ \text{Find k-nearest neighbors and indices}}\\
    	{ $\Gamma.append(Index)$}\\
    	{ $Populate(i,r,Index)$}
	}
	{$p\gets$ a number between 1 and k}\\
      {$\alpha\gets$ a number between 0 and 1}\\
        \For{$j \gets 1, 2,..., f$}{
    	{ $\Lambda = S[j]-S^{\prime}[j];$}\\
    	 { $S_{new}[j] = S[j]+\alpha * \Lambda;$}\\

	}
		
    {Return $S_{new}$}
    \caption{Pseudo-code for handling class imbalanced issue}
    \label{imbalancedIssue}
    \end{algorithm}

After all operations explained above are done, three more steps are required before the pre-processed data is given to a supervised algorithm.

\begin{enumerate}
  \item Tokenisation: each review is broken into words (called tokens).
  \item Vectorisation: each review is converted into a numeric representation (called corpus).
  \item Transformation: each review is transformed into one row (including 0 or 1) where 1 is the word in the corpus corresponding to that column appearing in that review.
\end{enumerate}

\subsection{Supervised Learning}
Text normalization was used to convert text into more convenient, standard forms. Tokenisation was used to separate words from running text. Each review has a rate\_value between 1-5. Any rate\_value of 1, 2, and 3 is considered a negative review; 4 and 5 are considered a positive review. Thus, there are two classes in this work, as the methods implemented are binary classification models. Then tokenized words are converted into a numeric representation, a process known as vectorization. After the data is processed, two approaches were applied: unigrams and n-grams. An $n$-gram is a contiguous sequence of $n$ words collected from our reviews. When $n$ is equal to $1$, it refers to as a unigram. Their corresponding models are probabilistic language models for predicting every word's ratio (in a unigram approach) or sequence of words (in an $n$-gram approach). After all the described operations are done, the pre-processed data is trained on a supervised ML method, i.e., SVM.

\begin{figure}
  \includegraphics[width=\linewidth]{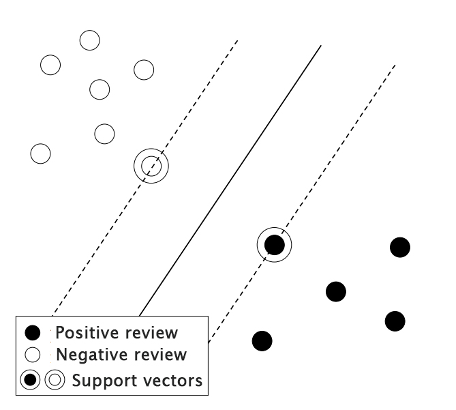}
  \caption{The support vectors given positive and negative classes.}
  \label{SVMFig}
\end{figure}

SVM is incorporated as a discriminative classifier for document categorization in this work. As explained in the prior section, it is less sensitive to the class imbalanced problems. This technique is based on the Structural Risk Minimisation principle. SVM's task is to learn and generalize an input-output mapping by finding separation between hyperplanes defined by classes of data. In our case, the set of reviews is the algorithm input, and their respective labels are the output. SVM searches for a separating hyperplane, which separates positive and negative reviews from each other with maximal margin; in other words, the distance of the decision surface and the closest review is maximal (Fig. \ref{SVMFig}).

Let $(x_1,y_1), (x_2,y_2) , ..., (x_n,y_n)$, $y_i \in \{positive, negative\}$ be our training observations. The SVM classifier is implemented by solving the following optimisation problem:

    \begin{equation}
maximise     \sum_{i=1}^{n} \mu_i - \frac{1}{2} \sum_{i,j=1}^{n}\mu_i\mu_jy_iy_j\phi(x_i,x_j)
  \end{equation}
    \begin{equation}
    \begin{multlined}
f(x)=  \sum_{i=1}^{n}y_i \mu_i \phi(x_i,x_j)+\xi \\
\forall i:0\leq \mu_i \leq C  \text{     }\text{     } and \text{  }\sum_{i,j=1}^{n}\mu_iy_i=0
\end{multlined}
  \end{equation}

where $\phi$ is a pisa kernel function, $\mu$ is a weight value, $\xi$ is a threshold and $C$ is a misclassification cost. The algorithm offers an optimal hyperplane, which is a decision boundary between the two classes.
 
\section{Results}\label{Results}
Supervised machine learning approaches are about conducting algorithms that precisely project a given input features to an output space. Each of these methods operates in two stages. First, an algorithm is trained based on a training dataset. Then, the algorithm is evaluated over various metrics based on a test dataset. Splitting the dataset is essential for an unbiased evaluation of prediction performance. Hence, the dataset used in this work was divided into two subsets. The testing dataset includes $3000$ reviews, consisting of $2714$ positive and $286$ negative comments. As explained above, this dataset is used for the evaluation of all models implemented in this work. It should be mentioned that the imbalanced handling phase is implemented when the training data is fitted.

Given the above discussion, the training set was applied to train models. Computational analyses were implemented based on two scenarios, i.e., a traditional approach and the model proposed in this work. Both scenarios include all the data handling steps explained earlier, i.e., data pre-processing and supervised learning. However, our proposed model includes an additional imbalanced handling phase described in the previous section. As far as the first scenario is concerned, various supervised algorithms, including Deep Neural Network (DNN), Recurrent Neural Network (RNN), Quadratic Discriminant Analysis (QDA), and Random Forest (RF), are tested and their results are compared with the proposed model.  

Given the extracted features, all the mentioned algorithms were fitted. After training all models, we have evaluated them to verify their applicability. Understanding how a model performs is essential to the use and development of text classification methods. To do so, the area under the Receiver Operating Characteristics (ROC) curves are used for comparing the accuracy of algorithms. These curves reveal a trade-off between the true positive rate and the false positive rate. The evaluation metric is based on a confusion matrix that comprises true positives (TP), false positives (FP), false negatives (FN), and true negatives (TN). The significance of these four elements may vary based on the classification application. In this work, the fraction of correct predictions overall predictions is considered.

    \begin{equation}
accuracy=\frac{TP+TN}{TP+FP+FN+TN}  
  \end{equation}

\begin{figure}
  \includegraphics[width=\linewidth]{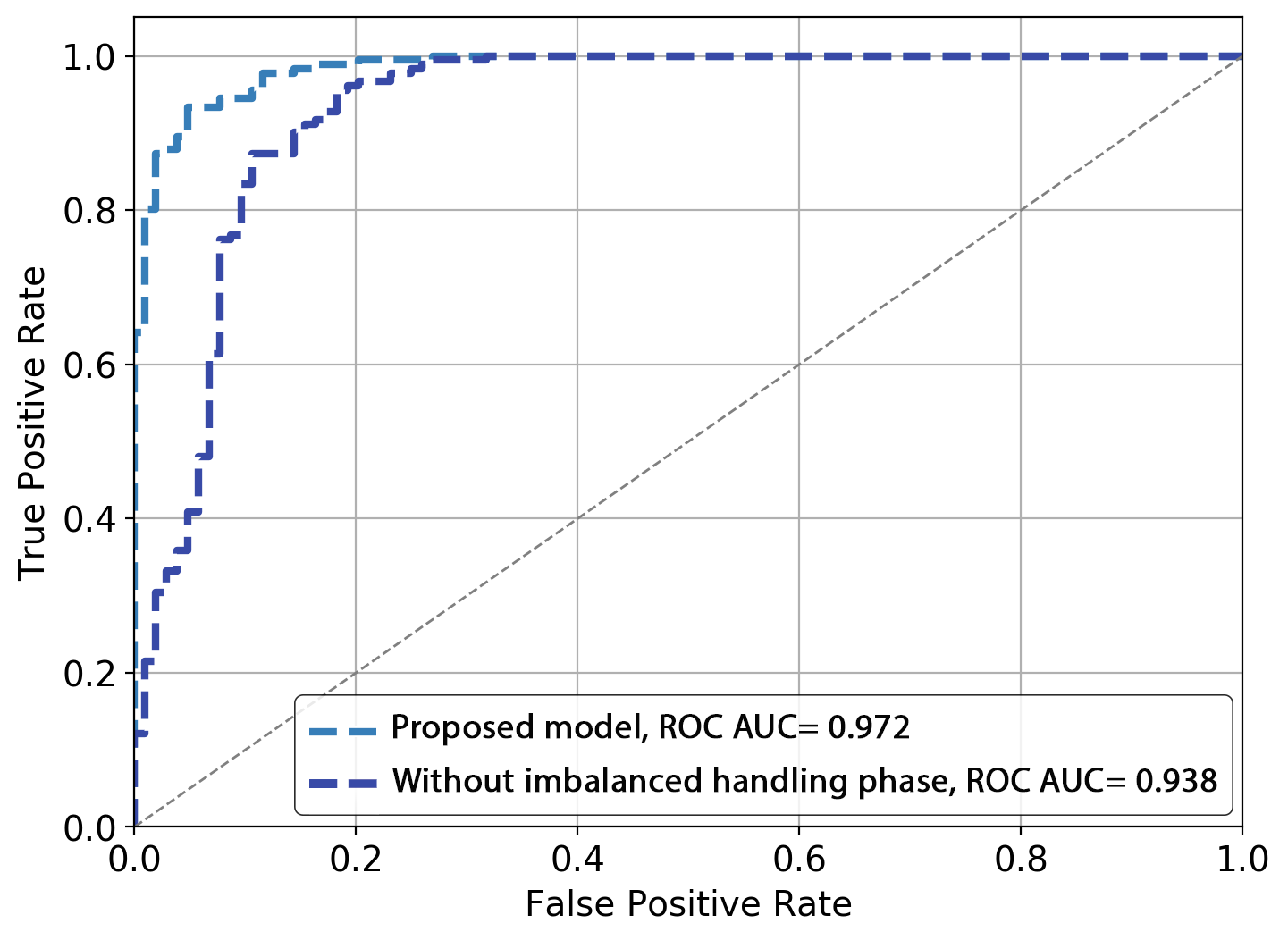}
  \caption{Proposed model ROC curves with and without imbalanced handling phase.}
  \label{proposedAUC}
\end{figure}

\begin{figure}
  \includegraphics[width=\linewidth]{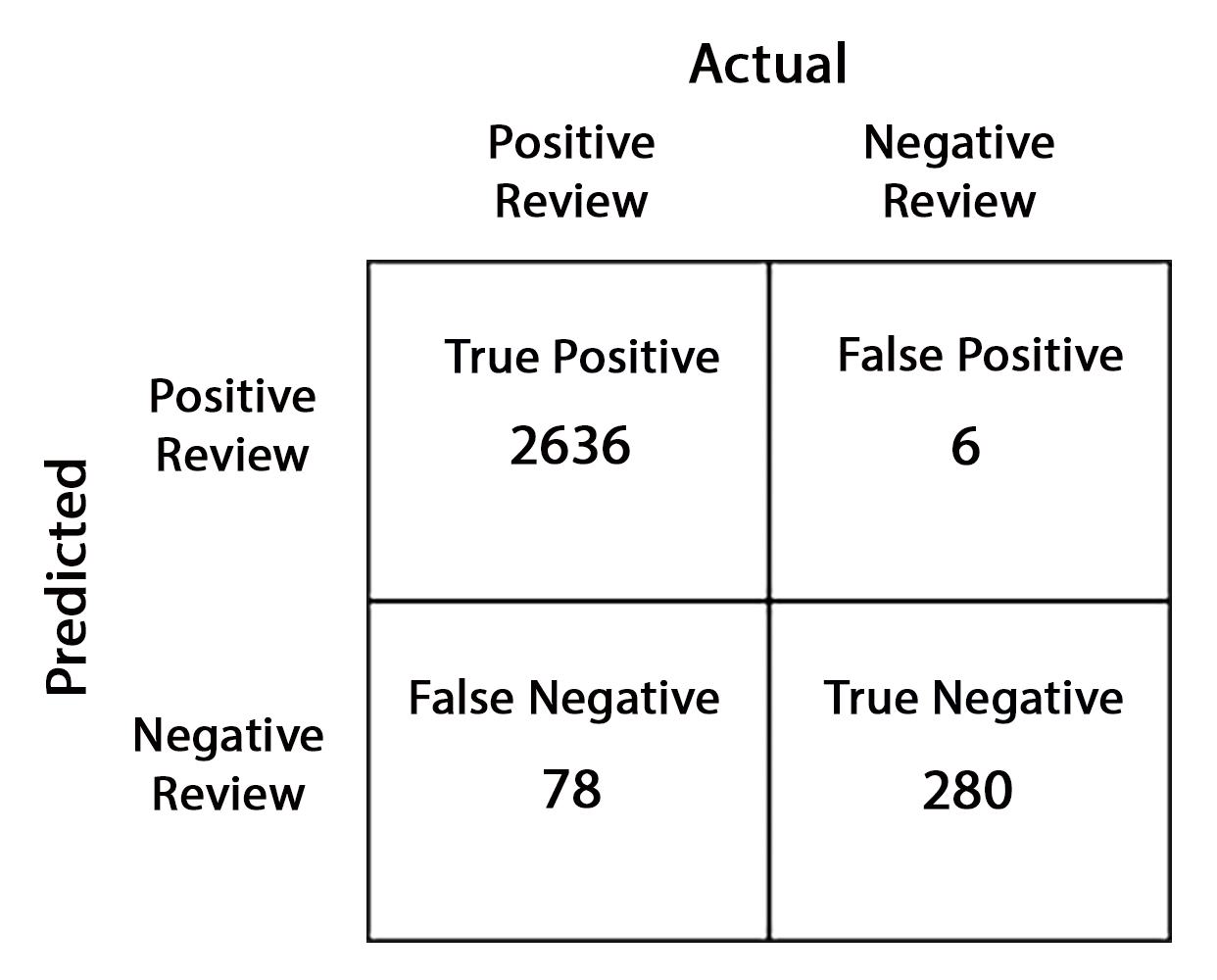}
  \caption{Confusion matrix for the positive and negative classes.}
  \label{cmatrix}
\end{figure}

Fig. \ref{proposedAUC} illustrates the ROC curve given the proposed model in this paper with and without the imbalanced handling phase. As shown, the model's predictive accuracy, assessed using the area under the curve (AUC), is over 97\%. The confusion matrix is also presented in Fig. \ref{cmatrix}.

\begin{figure}
  \includegraphics[width=\linewidth]{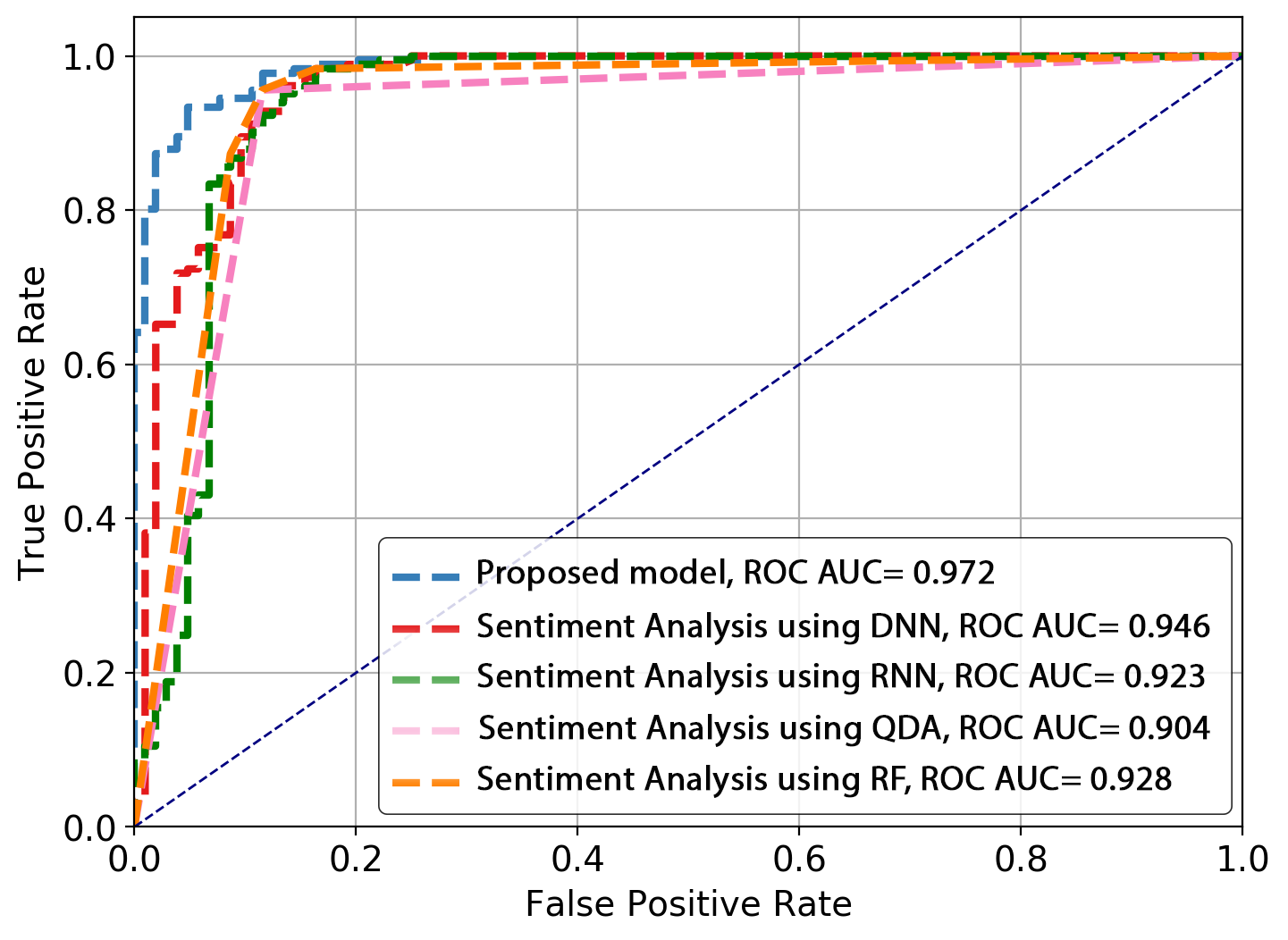}
  \caption{ROC curves for different approaches.}
  \label{AUCModels}
\end{figure}

As stated, the proposed model has been experimentally validated and compared with four different approaches. Their corresponding performances have been evaluated according to their classification accuracies. The results are depicted in Fig \ref{AUCModels}. The ability of each method to accurately predict the correct class is measured and expressed as a percentage. ROC curves have been used to determine the predictive performance of the examined classification algorithms. The area under a ROC has been considered as an evaluation criterion to select the best classification algorithm. When the area under the curve is approaching 1, it indicates that the classification was carried out correctly. We have also tested three more metrics, i.e., Precision, Recall, and $F1$-Score (Table \ref{ModelsMetrics}). The Recall metric is the measure of the correctly predicted positive reviews from all the actual positive ones (Recall $= \frac{TP}{TP+FN}$). Hence, it is a good indicator for evaluating models (given the cost of False Negatives) dealing with the imbalanced class issue.

All experimental results show that our proposed model is superior to those tested. The additional imbalanced handling phase incorporated improves the fit of the model.\\

\begin{table}
\caption{Comparisons of different classification metrics given 5 tested approaches.}
  \label{ModelsMetrics} 
\small
\begin{tabular}{l*{6}{c}r}
Models & Precision & Recall & F1 score  \\
\hline
Proposed approach & 0.971 & 0.997 & 0.983   \\
DNN-based model&    0.946 & 0.993 & 0.968 &  \\
RNN-based model&  0.92 & 0.994 & 0.955   \\
QDA-based model&  0.901 & 0.991 & 0.942   \\
RF-based model&  0.927 & 0.992 & 0.957   \\
\end{tabular}
\end{table}

\section{Discussion}\label{Discussion}
TripAdvisor reviews reveal a treasure trove of global, comparative data. To date, no other crowdsourced data was widespread enough to allow for such comparison between green spaces, both within the city and beyond. The most common criticism of observational approaches was time and cost expenditures spent on repeat measurements at the same locations (\cite{Cohen2009Recreational}). Like TripAdvisor, Twitter-based methods overcame this hurdle; tweets can be captured easily and frequently, offering greater measurement and (even longitudinal) analysis opportunities, saving time and costs (\cite{Roberts2017Geo}). However, TripAdvisor, for a city's most popular parks, offers more data. The TripAdvisor reviews collected for this study averaged out to be 80 reviews/month/park for St. Stephen's Green. Roberts (\cite{Roberts2017Geo}) study collected about 11 tweets/month/park for her study area. Furthermore, tweets and reviews are not the same; while a tweet can be about UGS, a TripAdvisor review explicitly asks for a reviewer's experience (i.e., a person's sentiment).

The elderly population, who show lower levels of engagement with technology in general, are specifically overlooked in crowdsourced data-based research (\cite{Chen2017Delineating}). This is especially concerning as UGS are intended to be a shared public space for all ages. Roberts also reports Twitter data lack demographic information about Twitter users, such as their age, occupation, or ethnicity (\cite{Roberts2017Geo}). These parameters, although not crucial to determine opinions, are useful for further examination of where particular attitudes may originate. Research to validate these demographic claims is limited, and studies comparing TripAdvisor with Twitter’s user base are non-existent. Although some groups remain over- and/or underrepresented on TripAdvisor, there is an option to collect some demographic information such as gender, nationality, and age. We acknowledge the bias most crowdsourced data has and understand it is both a contested and fertile research area.

TripAdvisor enabled us to utilize the abundance of open-source reviews. The accessibility of the reviews makes the proposed method highly scalable—especially for popular parks. In Dublin, 33 of its 50 parks are listed on TripAdvisor. However, besides the most popular St. Stephen's Green (16,613 reviews), Phoenix Park (4,753 reviews), and St. Anne’s Park (244 reviews), the remaining 30 parks have 1-66 reviews. Worldwide, thousands of UGS, from large to small, are listed on TripAdvisor. However, Dublin follows a similar pattern as other cities, where the most popular parks have significant reviews, and the lesser common parks have significantly fewer reviews. Therefore, we suggest the proposed method to be used only on a city's most popular parks, as a proxy for UGS in cities, and then compare UGS between cities worldwide. By leveraging machine learning techniques for opinion mining and text classification, hundreds of thousands of opinions previously overlooked can now be heard in an effort to improve these vulnerable public spaces.

\section{Conclusions}\label{conclusion}
Research to inform both policy and design of UGS is critical to protect these vulnerable areas while simultaneously ensuring access to the potential health and well-being benefits these spaces provide. Green spaces play a pivotal role across all aspects of city life, and as cities densify, the importance of accurately and effectively measuring the quality of UGS has never been greater. 

This paper presents an experiment's results to use NLP to extract citizen opinion on the quality of UGS, a highly novel application of automatic text classification on TripAdvisor reviews. The results indicate that the proposed method performs better, at $97$ accuracy, which is better than other approaches tested in this work.

Citizens, collectively, can enact meaningful change by acting as "ground agents" and providing valuable insights directly from the front lines. In this regard, citizens' insights are a goldmine of data that organizations can use to make their cities smarter. The results presented in this paper hold the potential to harness those opinions and give urban planners and local authorities greater choice to identify, analyze, and improve the sentiment behind specific UGS, and allow UGS comparisons between cities worldwide.

\bibliographystyle{cas-model2-names}

\bibliography{cas-refs}

\begin{thebibliography}{48}
\expandafter\ifx\csname natexlab\endcsname\relax\def\natexlab#1{#1}\fi
\providecommand{\url}[1]{\texttt{#1}}
\providecommand{\href}[2]{#2}
\providecommand{\path}[1]{#1}
\providecommand{\DOIprefix}{doi:}
\providecommand{\ArXivprefix}{arXiv:}
\providecommand{\URLprefix}{URL: }
\providecommand{\Pubmedprefix}{pmid:}
\providecommand{\doi}[1]{\href{http://dx.doi.org/#1}{\path{#1}}}
\providecommand{\Pubmed}[1]{\href{pmid:#1}{\path{#1}}}
\providecommand{\bibinfo}[2]{#2}
\ifx\xfnm\relax \def\xfnm[#1]{\unskip,\space#1}\fi
%Type = Article
\bibitem[{{Al Amrani} et~al.(2018){Al Amrani}, Lazaar and {El
  Kadiri}}]{ALAMRANI2018511}
\bibinfo{author}{{Al Amrani}, Y.}, \bibinfo{author}{Lazaar, M.},
  \bibinfo{author}{{El Kadiri}, K.E.}, \bibinfo{year}{2018}.
\newblock \bibinfo{title}{Random forest and support vector machine based hybrid
  approach to sentiment analysis}.
\newblock \bibinfo{journal}{Procedia Computer Science} \bibinfo{volume}{127},
  \bibinfo{pages}{511 -- 520}.
\newblock \DOIprefix\doi{https://doi.org/10.1016/j.procs.2018.01.150}.
  \bibinfo{note}{pROCEEDINGS OF THE FIRST INTERNATIONAL CONFERENCE ON
  INTELLIGENT COMPUTING IN DATA SCIENCES, ICDS2017}.
%Type = Article
\bibitem[{Barton and Rogerson(2017)}]{Barton2017importance}
\bibinfo{author}{Barton, J.}, \bibinfo{author}{Rogerson, M.},
  \bibinfo{year}{2017}.
\newblock \bibinfo{title}{The importance of greenspace for mental health.}
\newblock \bibinfo{journal}{BJPsych Int.} \bibinfo{volume}{14},
  \bibinfo{pages}{79--81}.
\newblock \DOIprefix\doi{10.1192/s2056474000002051}.
%Type = Misc
\bibitem[{Blumenthal(2014)}]{BlumenthalM2014}
\bibinfo{author}{Blumenthal, M.}, \bibinfo{year}{2014}.
\newblock \bibinfo{title}{Reviewer demographics-facebook has more women, yelp
  has more men.}
\newblock
  \bibinfo{howpublished}{http://blumenthals.com/blog/2014/07/31/reviewer-demographics-facebook-has-more-women-yelp-has-more-men/}.
%Type = Article
\bibitem[{Chen et~al.()Chen, Lee and Kirk}]{Chen2017Engaging}
\bibinfo{author}{Chen, Y.}, \bibinfo{author}{Lee, B.}, \bibinfo{author}{Kirk,
  R.M.}, .
\newblock \bibinfo{title}{Internet use among older adults.}
\newblock \bibinfo{journal}{Engaging Older Adults with Modern Technology} ,
  \bibinfo{pages}{124–141}.
%Type = Article
\bibitem[{Chen et~al.(2017)Chen, Liu, Li, Liu, Yao, Hu, Xu and
  Pei}]{Chen2017Delineating}
\bibinfo{author}{Chen, Y.}, \bibinfo{author}{Liu, X.}, \bibinfo{author}{Li,
  X.}, \bibinfo{author}{Liu, X.}, \bibinfo{author}{Yao, Y.},
  \bibinfo{author}{Hu, G.}, \bibinfo{author}{Xu, X.}, \bibinfo{author}{Pei,
  F.}, \bibinfo{year}{2017}.
\newblock \bibinfo{title}{Delineating urban functional areas with
  building-level social media data: A dynamic time warping (dtw) distance based
  k -medoids method.}
\newblock \bibinfo{journal}{Landscape and Urban Planning}
  \bibinfo{volume}{160}, \bibinfo{pages}{48–60}.
%Type = Article
\bibitem[{Cohen et~al.(2009)Cohen, Sehgal, Williamson, Marsh, Golinelli and
  McKenzie}]{Cohen2009Recreational}
\bibinfo{author}{Cohen, D.A.}, \bibinfo{author}{Sehgal, A.},
  \bibinfo{author}{Williamson, S.}, \bibinfo{author}{Marsh, T.},
  \bibinfo{author}{Golinelli, D.}, \bibinfo{author}{McKenzie, T.L.},
  \bibinfo{year}{2009}.
\newblock \bibinfo{title}{New recreational facilities for the young and the old
  in los angeles: Policy and programming implications.}
\newblock \bibinfo{journal}{Journal of Public Health Policy}
  \bibinfo{volume}{30}, \bibinfo{pages}{S248–S263}.
%Type = Article
\bibitem[{Colón-Ruiz and Segura-Bedmar(2020)}]{COLONRUIZ2020103539}
\bibinfo{author}{Colón-Ruiz, C.}, \bibinfo{author}{Segura-Bedmar, I.},
  \bibinfo{year}{2020}.
\newblock \bibinfo{title}{Comparing deep learning architectures for sentiment
  analysis on drug reviews}.
\newblock \bibinfo{journal}{Journal of Biomedical Informatics}
  \bibinfo{volume}{110}, \bibinfo{pages}{103539}.
\newblock \DOIprefix\doi{https://doi.org/10.1016/j.jbi.2020.103539}.
%Type = Article
\bibitem[{Daniels et~al.(2018)Daniels, Zaunbrecher, Paas, Ottermanns, Ziefle
  and Rob-Nickoll}]{Daniels2018Assessment}
\bibinfo{author}{Daniels, B.}, \bibinfo{author}{Zaunbrecher, B.S.},
  \bibinfo{author}{Paas, B.}, \bibinfo{author}{Ottermanns, R.},
  \bibinfo{author}{Ziefle, M.}, \bibinfo{author}{Rob-Nickoll, M.},
  \bibinfo{year}{2018}.
\newblock \bibinfo{title}{Assessment of urban green space structures and their
  quality from a multidimensional perspective.}
\newblock \bibinfo{journal}{The Science of the Total Environment}
  \bibinfo{volume}{615}, \bibinfo{pages}{1364–1378}.
%Type = Article
\bibitem[{Dong(2020)}]{DONG2020103418}
\bibinfo{author}{Dong, J.}, \bibinfo{year}{2020}.
\newblock \bibinfo{title}{Financial investor sentiment analysis based on fpga
  and convolutional neural network}.
\newblock \bibinfo{journal}{Microprocessors and Microsystems} ,
  \bibinfo{pages}{103418}\DOIprefix\doi{https://doi.org/10.1016/j.micpro.2020.103418}.
%Type = Article
\bibitem[{Douglas et~al.(2017)Douglas, Lennon and Scott}]{Douglas2017Green}
\bibinfo{author}{Douglas, O.}, \bibinfo{author}{Lennon, M.},
  \bibinfo{author}{Scott, M.}, \bibinfo{year}{2017}.
\newblock \bibinfo{title}{Green space benefits for health and well-being: A
  life-course approach for urban planning, design and management.}
\newblock \bibinfo{journal}{Cities} \bibinfo{volume}{66},
  \bibinfo{pages}{53–62}.
%Type = Article
\bibitem[{Fang et~al.(2020)Fang, McNeil, Warner, Dahle and
  Eutsler}]{Fang2020Street}
\bibinfo{author}{Fang, F.}, \bibinfo{author}{McNeil, B.},
  \bibinfo{author}{Warner, T.}, \bibinfo{author}{Dahle, G.},
  \bibinfo{author}{Eutsler, E.}, \bibinfo{year}{2020}.
\newblock \bibinfo{title}{Street tree health from space? an evaluation using
  worldview-3 data and the washington dc street tree spatial database.}
\newblock \bibinfo{journal}{Urban Forestry \& Urban Greening}
  \bibinfo{volume}{49}.
%Type = Article
\bibitem[{Fitri et~al.(2019)Fitri, Andreswari and Hasibuan}]{FITRI2019765}
\bibinfo{author}{Fitri, V.A.}, \bibinfo{author}{Andreswari, R.},
  \bibinfo{author}{Hasibuan, M.A.}, \bibinfo{year}{2019}.
\newblock \bibinfo{title}{Sentiment analysis of social media twitter with case
  of anti-lgbt campaign in indonesia using naïve bayes, decision tree, and
  random forest algorithm}.
\newblock \bibinfo{journal}{Procedia Computer Science} \bibinfo{volume}{161},
  \bibinfo{pages}{765 -- 772}.
\newblock \DOIprefix\doi{https://doi.org/10.1016/j.procs.2019.11.181}.
  \bibinfo{note}{the Fifth Information Systems International Conference, 23-24
  July 2019, Surabaya, Indonesia}.
%Type = Article
\bibitem[{Fuller and Gaston(2009)}]{Fuller2009Green}
\bibinfo{author}{Fuller, R.A.}, \bibinfo{author}{Gaston, K.J.},
  \bibinfo{year}{2009}.
\newblock \bibinfo{title}{The scaling of green space coverage in european
  cities.}
\newblock \bibinfo{journal}{Biology Letters} \bibinfo{volume}{5},
  \bibinfo{pages}{352–355}.
%Type = Article
\bibitem[{Galle et~al.(2003)Galle, Nitoslawski and Pilla}]{Galle2019Internet}
\bibinfo{author}{Galle, N.J.}, \bibinfo{author}{Nitoslawski, S.A.},
  \bibinfo{author}{Pilla, F.}, \bibinfo{year}{2003}.
\newblock \bibinfo{title}{The internet of nature: How taking nature online can
  shape urban ecosystems.}
\newblock \bibinfo{journal}{The Anthropocene Review} \bibinfo{volume}{6},
  \bibinfo{pages}{279–287}.
\newblock \DOIprefix\doi{https://doi.org/10.1177/2053019619877103}.
%Type = Inproceedings
\bibitem[{Garcia-Barriocanal et~al.(2010)Garcia-Barriocanal, Sicilia and
  Korfiatis}]{Garcia2010}
\bibinfo{author}{Garcia-Barriocanal, E.}, \bibinfo{author}{Sicilia, M.},
  \bibinfo{author}{Korfiatis, N.}, \bibinfo{year}{2010}.
\newblock \bibinfo{title}{Exploring hotel service quality experience indicators
  in user-generated content: a case using tripadvisor data.}, in:
  \bibinfo{booktitle}{Mediterranean Conference on Information Systems}, p.
  \bibinfo{pages}{200–205}.
%Type = Article
\bibitem[{{Ghahramani} et~al.(2020){Ghahramani}, {Qiao}, {Zhou}, {O'Hagan} and
  {Sweeney}}]{Ghahramani2020AI}
\bibinfo{author}{{Ghahramani}, M.}, \bibinfo{author}{{Qiao}, Y.},
  \bibinfo{author}{{Zhou}, M.C.}, \bibinfo{author}{{O'Hagan}, A.},
  \bibinfo{author}{{Sweeney}, J.}, \bibinfo{year}{2020}.
\newblock \bibinfo{title}{Ai-based modeling and data-driven evaluation for
  smart manufacturing processes}.
\newblock \bibinfo{journal}{IEEE/CAA Journal of Automatica Sinica}
  \bibinfo{volume}{7}, \bibinfo{pages}{1026--1037}.
\newblock \DOIprefix\doi{10.1109/JAS.2020.1003114}.
%Type = Article
\bibitem[{Ghahramani et~al.(2020)Ghahramani, Zhou and
  Wang}]{Ghahramani2020Urban}
\bibinfo{author}{Ghahramani, M.}, \bibinfo{author}{Zhou, M.},
  \bibinfo{author}{Wang, G.}, \bibinfo{year}{2020}.
\newblock \bibinfo{title}{Urban sensing based on mobile phone data: approaches,
  applications, and challenges}.
\newblock \bibinfo{journal}{IEEE/CAA Journal of Automatica Sinica}
  \bibinfo{volume}{7}, \bibinfo{pages}{627--637}.
%Type = Article
\bibitem[{Gidlow et~al.(2012)Gidlow, Ellis and Bostock}]{Gidlow2012Development}
\bibinfo{author}{Gidlow, C.J.}, \bibinfo{author}{Ellis, N.J.},
  \bibinfo{author}{Bostock, S.}, \bibinfo{year}{2012}.
\newblock \bibinfo{title}{Development of the neighbourhood green space tool
  (ngst).}
\newblock \bibinfo{journal}{Landscape and Urban Planning}
  \bibinfo{volume}{106}, \bibinfo{pages}{347–358}.
%Type = Article
\bibitem[{Giezen et~al.(2018)Giezen, Balikci and Arundel}]{Giezen2018Using}
\bibinfo{author}{Giezen, M.}, \bibinfo{author}{Balikci, S.},
  \bibinfo{author}{Arundel, R.}, \bibinfo{year}{2018}.
\newblock \bibinfo{title}{Using remote sensing to analyse net land-use change
  from conflicting sustainability policies: The case of amsterdam.}
\newblock \bibinfo{journal}{ISPRS International Journal of Geo-Information}
  \bibinfo{volume}{7}.
\newblock \DOIprefix\doi{https://doi.org/10.3390/ijgi7090381}.
%Type = Misc
\bibitem[{Girling and Kellett(2005)}]{Girling2005Skinny}
\bibinfo{author}{Girling, C.}, \bibinfo{author}{Kellett, R.},
  \bibinfo{year}{2005}.
\newblock \bibinfo{title}{Skinny streets and green neighborhoods: Design for
  environment and community.}
\newblock \bibinfo{howpublished}{Island Press.}
%Type = Article
\bibitem[{Grahn and Stigsdotter(2010)}]{Grahn2010relation}
\bibinfo{author}{Grahn, P.}, \bibinfo{author}{Stigsdotter, U.K.},
  \bibinfo{year}{2010}.
\newblock \bibinfo{title}{The relation between perceived sensory dimensions of
  urban green space and stress restoration.}
\newblock \bibinfo{journal}{Landscape and Urban Planning} \bibinfo{volume}{94},
  \bibinfo{pages}{264–275}.
%Type = Misc
\bibitem[{Gretzel(2007)}]{GretzelU}
\bibinfo{author}{Gretzel, U.}, \bibinfo{year}{2007}.
\newblock \bibinfo{title}{Online travel review study: Role and impact of online
  travel reviews.}
\newblock \bibinfo{howpublished}{Laboratory for Intelligent Systems in
  Tourism}.
%Type = Article
\bibitem[{Haaland and van~den Bosch(2015)}]{Haaland2015Challenges}
\bibinfo{author}{Haaland, C.}, \bibinfo{author}{van~den Bosch, C.K.},
  \bibinfo{year}{2015}.
\newblock \bibinfo{title}{Challenges and strategies for urban green-space
  planning in cities undergoing densification: A review.}
\newblock \bibinfo{journal}{Urban Forestry \& Urban Greening}
  \bibinfo{volume}{14}, \bibinfo{pages}{760–771}.
\newblock \DOIprefix\doi{https://doi.org/10.1016/j.ufug.2015.07.009}.
%Type = Article
\bibitem[{Hu et~al.(2020)Hu, O’Hagan, Sweeney and
  Ghahramani}]{Ghahramani2020Spatial}
\bibinfo{author}{Hu, S.}, \bibinfo{author}{O’Hagan, A.},
  \bibinfo{author}{Sweeney, J.}, \bibinfo{author}{Ghahramani, M.},
  \bibinfo{year}{2020}.
\newblock \bibinfo{title}{A spatial machine learning model for analysing
  customers’ lapse behaviour in life insurance}.
\newblock \bibinfo{journal}{Annals of Actuarial Science}
  \DOIprefix\doi{https://doi.org/10.1017/S1748499520000329}.
%Type = Article
\bibitem[{Hur et~al.(2010)Hur, Nasar and Chun}]{Hur2010Neighborhood}
\bibinfo{author}{Hur, M.}, \bibinfo{author}{Nasar, J.L.},
  \bibinfo{author}{Chun, B.}, \bibinfo{year}{2010}.
\newblock \bibinfo{title}{Neighborhood satisfaction, physical and perceived
  naturalness and openness.}
\newblock \bibinfo{journal}{Journal of Environmental Psychology}
  \bibinfo{volume}{30}, \bibinfo{pages}{52–59}.
\newblock \DOIprefix\doi{https://doi.org/10.1016/j.jenvp.2009.05.005}.
%Type = Article
\bibitem[{J. and Verburg P.~H.(2018)}]{Van2018Aesthetic}
\bibinfo{author}{J., V.Z.B.T.S.C.}, \bibinfo{author}{Verburg P.~H., T.K.F.},
  \bibinfo{year}{2018}.
\newblock \bibinfo{title}{Aesthetic appreciation of the cultural landscape
  through social media: An analysis of revealed preference in the dutch river
  landscape.}
\newblock \bibinfo{journal}{Landscape and Urban Planning}
  \bibinfo{volume}{177}, \bibinfo{pages}{128–137}.
%Type = Article
\bibitem[{{Jin} et~al.(2020a){Jin}, {Wu}, {Ma}, {Yan} and
  {Mo}}]{Jin2020Learning}
\bibinfo{author}{{Jin}, N.}, \bibinfo{author}{{Wu}, J.}, \bibinfo{author}{{Ma},
  X.}, \bibinfo{author}{{Yan}, K.}, \bibinfo{author}{{Mo}, Y.},
  \bibinfo{year}{2020}a.
\newblock \bibinfo{title}{Multi-task learning model based on multi-scale cnn
  and lstm for sentiment classification}.
\newblock \bibinfo{journal}{IEEE Access} \bibinfo{volume}{8},
  \bibinfo{pages}{77060--77072}.
\newblock \DOIprefix\doi{10.1109/ACCESS.2020.2989428}.
%Type = Article
\bibitem[{{Jin} et~al.(2020b){Jin}, {Wu}, {Ma}, {Yan} and {Mo}}]{9076160}
\bibinfo{author}{{Jin}, N.}, \bibinfo{author}{{Wu}, J.}, \bibinfo{author}{{Ma},
  X.}, \bibinfo{author}{{Yan}, K.}, \bibinfo{author}{{Mo}, Y.},
  \bibinfo{year}{2020}b.
\newblock \bibinfo{title}{Multi-task learning model based on multi-scale cnn
  and lstm for sentiment classification}.
\newblock \bibinfo{journal}{IEEE Access} \bibinfo{volume}{8},
  \bibinfo{pages}{77060--77072}.
\newblock \DOIprefix\doi{10.1109/ACCESS.2020.2989428}.
%Type = Article
\bibitem[{Kourtit et~al.(2019)Kourtit, Nijkamp and
  Romao.}]{Kourtit2019Cultural}
\bibinfo{author}{Kourtit}, \bibinfo{author}{Nijkamp}, \bibinfo{author}{Romao.},
  \bibinfo{year}{2019}.
\newblock \bibinfo{title}{Cultural heritage appraisal by visitors to global
  cities: The use of social media and urban analytics in urban buzz research.}
\newblock \bibinfo{journal}{Sustainability} \bibinfo{volume}{11},
  \bibinfo{pages}{3470}.
\newblock \DOIprefix\doi{https://doi.org/10.3390/su11123470}.
%Type = Article
\bibitem[{Ma et~al.(2018)Ma, Cheng and Hsiao}]{Ma2018Sentiment}
\bibinfo{author}{Ma, E.}, \bibinfo{author}{Cheng, M.}, \bibinfo{author}{Hsiao,
  A.}, \bibinfo{year}{2018}.
\newblock \bibinfo{title}{Sentiment analysis – a review and agenda for future
  research in hospitality contexts.}
\newblock \bibinfo{journal}{International Journal of Contemporary Hospitality
  Management} \bibinfo{volume}{30}, \bibinfo{pages}{3287–3308}.
\newblock \DOIprefix\doi{https://doi.org/10.1108/ijchm-10-2017-0704}.
%Type = Article
\bibitem[{Nasi et~al.(2018)Nasi, Honkavaara, Blomqvist, Lyytikäinen-Saarenmaa,
  Hakala, Viljanen and Holopainen}]{Nasi2018Remote}
\bibinfo{author}{Nasi, R.}, \bibinfo{author}{Honkavaara, E.},
  \bibinfo{author}{Blomqvist, M.}, \bibinfo{author}{Lyytikäinen-Saarenmaa,
  P.}, \bibinfo{author}{Hakala, T.}, \bibinfo{author}{Viljanen, N.},
  \bibinfo{author}{Holopainen, M.}, \bibinfo{year}{2018}.
\newblock \bibinfo{title}{Remote sensing of bark beetle damage in urban forests
  at individual tree level using a novel hyperspectral camera from uav and
  aircraft.}
\newblock \bibinfo{journal}{Urban Forestry \& Urban Greening}
  \bibinfo{volume}{30}, \bibinfo{pages}{72--83}.
%Type = Article
\bibitem[{Nitoslawski et~al.(2019)Nitoslawski, Galle, Van Den~Bosch and
  Steenberg}]{Nitoslawski2018Smarter}
\bibinfo{author}{Nitoslawski, S.A.}, \bibinfo{author}{Galle, N.J.},
  \bibinfo{author}{Van Den~Bosch, C.K.}, \bibinfo{author}{Steenberg, J.W.N.},
  \bibinfo{year}{2019}.
\newblock \bibinfo{title}{Smarter ecosystems for smarter cities? a review of
  trends, technologies, and turning points for smart urban forestry.}
\newblock \bibinfo{journal}{Sustainable Cities and Society}
  \bibinfo{volume}{51}, \bibinfo{pages}{101770}.
\newblock \DOIprefix\doi{https://doi.org/10.1016/j.scs.2019.101770}.
%Type = Article
\bibitem[{Nowak and Greenfield(2018)}]{Nowak2018Declining}
\bibinfo{author}{Nowak, D.J.}, \bibinfo{author}{Greenfield, E.J.},
  \bibinfo{year}{2018}.
\newblock \bibinfo{title}{Declining urban and community tree cover in the
  united states.}
\newblock \bibinfo{journal}{Urban Forestry \& Urban Greening}
  \bibinfo{volume}{32}, \bibinfo{pages}{32–55}.
\newblock \DOIprefix\doi{https://doi.org/10.1016/j.ufug.2018.03.006}.
%Type = Article
\bibitem[{Pak et~al.(2017)Pak, Chua and Moere}]{Pak2017FixMyStreet}
\bibinfo{author}{Pak, B.}, \bibinfo{author}{Chua, A.}, \bibinfo{author}{Moere,
  A.V.}, \bibinfo{year}{2017}.
\newblock \bibinfo{title}{Fixmystreet brussels: Socio-demographic inequality in
  crowdsourced civic participation.}
\newblock \bibinfo{journal}{Journal of Urban Technology} \bibinfo{volume}{24},
  \bibinfo{pages}{65–87}.
%Type = Misc
\bibitem[{Pampel(2000)}]{Pampel}
\bibinfo{author}{Pampel, F.C.}, \bibinfo{year}{2000}.
\newblock \bibinfo{title}{Logistic regression: A primer.}
\newblock \bibinfo{howpublished}{SAGE}.
%Type = Article
\bibitem[{Plunz et~al.(2019)Plunz, Zhou, Vintimilla, Mckeown, Yu, Uguccioni and
  Sutto}]{Plunz2019Twitter}
\bibinfo{author}{Plunz, R.A.}, \bibinfo{author}{Zhou, Y.},
  \bibinfo{author}{Vintimilla, M.I.C.}, \bibinfo{author}{Mckeown, K.},
  \bibinfo{author}{Yu, T.}, \bibinfo{author}{Uguccioni, L.},
  \bibinfo{author}{Sutto, M.P.}, \bibinfo{year}{2019}.
\newblock \bibinfo{title}{Twitter sentiment in new york city parks as measure
  of well-being.}
\newblock \bibinfo{journal}{Landscape and Urban Planning}
  \bibinfo{volume}{189}, \bibinfo{pages}{235–246}.
\newblock \DOIprefix\doi{https://doi.org/10.1016/j.landurbplan.2019.04.024}.
%Type = Article
\bibitem[{Rahimi et~al.(2018)Rahimi, Mottahedi and Liu}]{Rahimi2018Geography}
\bibinfo{author}{Rahimi, S.}, \bibinfo{author}{Mottahedi, S.},
  \bibinfo{author}{Liu, X.}, \bibinfo{year}{2018}.
\newblock \bibinfo{title}{The geography of taste: Using yelp to study urban
  culture.}
\newblock \bibinfo{journal}{ISPRS International Journal of Geo-Information}
  \bibinfo{volume}{7}, \bibinfo{pages}{376}.
\newblock \DOIprefix\doi{https://doi.org/10.3390/ijgi7090376}.
%Type = Article
\bibitem[{Rakhmanov(2020)}]{RAKHMANOV2020194}
\bibinfo{author}{Rakhmanov, O.}, \bibinfo{year}{2020}.
\newblock \bibinfo{title}{A comparative study on vectorization and
  classification techniques in sentiment analysis to classify student-lecturer
  comments}.
\newblock \bibinfo{journal}{Procedia Computer Science} \bibinfo{volume}{178},
  \bibinfo{pages}{194 -- 204}.
\newblock \DOIprefix\doi{https://doi.org/10.1016/j.procs.2020.11.021}.
  \bibinfo{note}{9th International Young Scientists Conference in Computational
  Science, YSC2020, 05-12 September 2020}.
%Type = Article
\bibitem[{Roberts et~al.(2017)Roberts, Sadler and Chapman}]{Roberts2017Geo}
\bibinfo{author}{Roberts, H.}, \bibinfo{author}{Sadler, J.},
  \bibinfo{author}{Chapman, L.}, \bibinfo{year}{2017}.
\newblock \bibinfo{title}{Using twitter to investigate seasonal variation in
  physical activity in urban green space.}
\newblock \bibinfo{journal}{Geo: Geography and Environment}
  \bibinfo{volume}{4}.
%Type = Article
\bibitem[{Roe et~al.(2018)Roe, Thompson, Aspinall, Brewer, Duff, Miller,
  Mitchell and Clow}]{Roe2018Green}
\bibinfo{author}{Roe, J.J.}, \bibinfo{author}{Thompson, C.W.},
  \bibinfo{author}{Aspinall, P.A.}, \bibinfo{author}{Brewer, M.J.},
  \bibinfo{author}{Duff, E.I.}, \bibinfo{author}{Miller, D.},
  \bibinfo{author}{Mitchell, R.}, \bibinfo{author}{Clow, A.},
  \bibinfo{year}{2018}.
\newblock \bibinfo{title}{Green space and stress: evidence from cortisol
  measures in deprived urban communities.}
\newblock \bibinfo{journal}{International Journal of Environmental Research and
  Public Health} \bibinfo{volume}{10}, \bibinfo{pages}{4086–4103}.
%Type = Article
\bibitem[{Rokach and Maimon(2005)}]{Rokach2005Top}
\bibinfo{author}{Rokach, L.}, \bibinfo{author}{Maimon, O.},
  \bibinfo{year}{2005}.
\newblock \bibinfo{title}{Top-down induction of decision trees classifiers—a
  survey.}
\newblock \bibinfo{journal}{IEEE Transactions on Systems, Man and Cybernetics,
  Part C (Applications and Reviews)} \bibinfo{volume}{35},
  \bibinfo{pages}{476–487}.
%Type = Article
\bibitem[{Seiferling et~al.(2017)Seiferling, Naik, Ratti and
  Proulx}]{Seiferling2017Green}
\bibinfo{author}{Seiferling, I.}, \bibinfo{author}{Naik, N.},
  \bibinfo{author}{Ratti, C.}, \bibinfo{author}{Proulx, R.},
  \bibinfo{year}{2017}.
\newblock \bibinfo{title}{Green streets- quantifying and mapping urban trees
  with street-level imagery and computer vision.}
\newblock \bibinfo{journal}{Landscape and Urban Planning}
  \bibinfo{volume}{165}, \bibinfo{pages}{93--101}.
%Type = Article
\bibitem[{Swanwick et~al.(2003)Swanwick, Dunnett and
  Woolley}]{Swanwick2003Nature}
\bibinfo{author}{Swanwick, C.}, \bibinfo{author}{Dunnett, N.},
  \bibinfo{author}{Woolley, H.}, \bibinfo{year}{2003}.
\newblock \bibinfo{title}{Nature, role and value of green space in towns and
  cities: An overview.}
\newblock \bibinfo{journal}{Built Environment} \bibinfo{volume}{29},
  \bibinfo{pages}{94–106}.
\newblock \DOIprefix\doi{https://doi.org/10.1016/j.cities.2017.03.011}.
%Type = Article
\bibitem[{{Tra} et~al.(2019){Tra}, {Duong} and {Kim}}]{8785940}
\bibinfo{author}{{Tra}, V.}, \bibinfo{author}{{Duong}, B.},
  \bibinfo{author}{{Kim}, J.}, \bibinfo{year}{2019}.
\newblock \bibinfo{title}{Improving diagnostic performance of a power
  transformer using an adaptive over-sampling method for imbalanced data}.
\newblock \bibinfo{journal}{IEEE Transactions on Dielectrics and Electrical
  Insulation} \bibinfo{volume}{26}, \bibinfo{pages}{1325--1333}.
\newblock \DOIprefix\doi{10.1109/TDEI.2019.008034}.
%Type = Article
\bibitem[{Yahav et~al.(2019)Yahav, Shehory and Schwartz}]{Yahav2019Comments}
\bibinfo{author}{Yahav, I.}, \bibinfo{author}{Shehory, O.},
  \bibinfo{author}{Schwartz, D.}, \bibinfo{year}{2019}.
\newblock \bibinfo{title}{Comments mining with tf-idf: The inherent bias and
  its removal.}
\newblock \bibinfo{journal}{IEEE Transactions on Knowledge and Data
  Engineering} \bibinfo{volume}{31}, \bibinfo{pages}{437–450}.
\newblock \DOIprefix\doi{https://doi.org/10.1109/tkde.2018.2840127}.
%Type = Article
\bibitem[{{Yang} et~al.(2018){Yang}, {Kuang}, {Zhang} and {Zhang}}]{8065074}
\bibinfo{author}{{Yang}, X.}, \bibinfo{author}{{Kuang}, Q.},
  \bibinfo{author}{{Zhang}, W.}, \bibinfo{author}{{Zhang}, G.},
  \bibinfo{year}{2018}.
\newblock \bibinfo{title}{Amdo: An over-sampling technique for multi-class
  imbalanced problems}.
\newblock \bibinfo{journal}{IEEE Transactions on Knowledge and Data
  Engineering} \bibinfo{volume}{30}, \bibinfo{pages}{1672--1685}.
\newblock \DOIprefix\doi{10.1109/TKDE.2017.2761347}.
%Type = Article
\bibitem[{Yao and Wang(2020)}]{YAO2020101522}
\bibinfo{author}{Yao, F.}, \bibinfo{author}{Wang, Y.}, \bibinfo{year}{2020}.
\newblock \bibinfo{title}{Domain-specific sentiment analysis for tweets during
  hurricanes (dssa-h): A domain-adversarial neural-network-based approach}.
\newblock \bibinfo{journal}{Computers, Environment and Urban Systems}
  \bibinfo{volume}{83}, \bibinfo{pages}{101522}.
\newblock \DOIprefix\doi{https://doi.org/10.1016/j.compenvurbsys.2020.101522}.
%Type = Article
\bibitem[{Zhang et~al.(2017)Zhang, Van~den Berg, Van~Dijk and
  Weitkamp}]{Zhang2017Quality}
\bibinfo{author}{Zhang, Y.}, \bibinfo{author}{Van~den Berg, A.E.},
  \bibinfo{author}{Van~Dijk, T.}, \bibinfo{author}{Weitkamp, G.},
  \bibinfo{year}{2017}.
\newblock \bibinfo{title}{Quality over quantity: Contribution of urban green
  space to neighborhood satisfaction.}
\newblock \bibinfo{journal}{International Journal of Environmental Research and
  Public Health} \bibinfo{volume}{14}.
\newblock \DOIprefix\doi{https://doi.org/10.3390/ijerph14050535}.

\end{thebibliography}


\end{document}